\documentclass[twocolumn,english,aps,letterpaper]{revtex4}
\usepackage{mathptmx}
\usepackage[T1]{fontenc}
\usepackage[latin9]{inputenc}
\usepackage[letterpaper]{geometry}
\usepackage{color}
\usepackage{amsmath}
\usepackage{graphicx}
\usepackage{amssymb}

\makeatletter
\@ifundefined{textcolor}{}
{%
 \definecolor{BLACK}{gray}{0}
 \definecolor{WHITE}{gray}{1}
 \definecolor{RED}{rgb}{1,0,0}
 \definecolor{GREEN}{rgb}{0,1,0}
 \definecolor{BLUE}{rgb}{0,0,1}
 \definecolor{CYAN}{cmyk}{1,0,0,0}
 \definecolor{MAGENTA}{cmyk}{0,1,0,0}
 \definecolor{YELLOW}{cmyk}{0,0,1,0}
 }

\makeatother

\usepackage{babel}

\begin{document}

\title{A possible Efimov trimer state in a 3-component lithium-6 mixture}

\author{Pascal Naidon}

\email{pascal@cat.phys.s.u-tokyo.ac.jp}

\author{Masahito Ueda}

\affiliation{ERATO Macroscopic Quantum Project, JST, Tokyo, 113-0033 Japan}

\affiliation{Department of Physics, University of Tokyo, 7-3-1 Hongo, Bunkyo-ku,
Tokyo 113-0033, Japan}
\begin{abstract}
We consider the Efimov trimer theory as a possible framework to explain
recently observed losses by inelastic three-body collisions in a three-hyperfine-component
ultracold mixture of lithium 6. Within this framework, these losses
would arise chiefly from the existence of an Efimov trimer bound state
below the continuum of free triplets of atoms, and the loss maxima
(at certain values of an applied magnetic field) would correspond
to zero-energy resonances where the trimer dissociates into three
free atoms. Our results show that such a trimer state is indeed possible
given the two-body scattering lengths in the three-component lithium
mixture, and gives rise to two zero-energy resonances. The locations
of these resonances appear to be consistent with observed losses.
\end{abstract}
\maketitle
\begin{figure}
\includegraphics[clip]{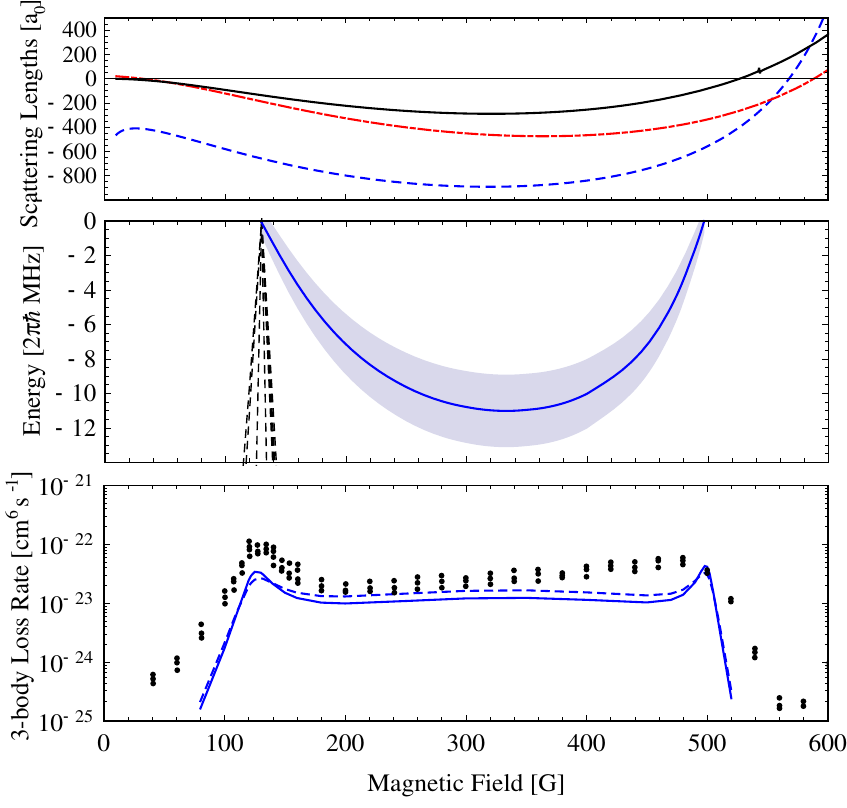}

\caption{\label{fig:Panels}\textbf{Top panel:} variation of the two-body scattering
lengths $a_{12}$, $a_{13}$, and $a_{23}$ for the lowest three hyperfine
components of lithium 6 as a function of magnetic field. These curves
were calculated by P.~S.~Julienne and taken from Ref.~\cite{rf:ottenstein}.\textbf{
Middle panel:} Energy of the Efimov trimer (solid curve) just below
the three-body threshold as a function of magnetic field with $\Lambda_{0}=(0.42\; a_{0})^{-1}$.
The shaded area corresponds to the width of the Efimov state, \emph{i.e.}
the imaginary part of its energy, for $\eta=0.115$. The estimated
energy of possible resonant trimers from all other spin channels with
the same total projection $m_{F}=-3/2$ is indicated by dashed curves.\textbf{
Bottom panel:} Experimental (dots, taken from Ref.~\cite{rf:ottenstein})
and theoretical (curves) three-body inelastic collision loss rate
coefficient as a function of magnetic field. The dashed curve is obtained
by adjusting the short-range loss parameter $\eta$ to fit the experimental
data ($\eta=0.157$), and the solid curve by adjusting it to fit the
shape of the left peak ($\eta=0.115$).}

\end{figure}

The experimental realisation of ultracold Fermi gases have let us
explore many fundamental aspects of few and many-body physics. In
particular, the study of mixtures of fermionic atoms in two different
spin components have led to the observation of superfluid paired phases
such as molecular Bose-Einstein condensates (BEC), Bardeen-Cooper-Schrieffer
(BCS) superfluids, and their crossover \cite{rf:regal,rf:zwierlein,rf:bartenstein,rf:bourdel,rf:partridge}.
Recently there has been some theoretical interest in Fermi systems
with three different spin components \cite{rf:bedaque,rf:zhai,rf:paananen,rf:blume,rf:cherng},
which can present analogies with colour superfluidity in QCD \cite{rf:rapp}.
Recent experiments \cite{rf:ottenstein,rf:huckans} have been performed
with ultracold mixtures of lithium 6 atoms prepared in the lowest
three hyperfine states $\vert1\rangle$, $\vert2\rangle$, and $\vert3\rangle$.
They indicated that when an external magnetic field is applied, strong
losses due to three-body inelastic collisions occur over a wide range
of magnetic field intensities. On the other hand, such losses are
not observed when only two of the three hyperfine components are mixed.
Therefore the observed inelastic collisions are related to the specific
scattering channel involving three atoms in the three different hyperfine
components. The magnitude of the inelastic collisions is characterised
by a loss rate coefficient $\mathcal{K}$, defined by the rate equation\[
\frac{dn_{i}}{dt}=-\mathcal{K}n_{i}n_{j}n_{k},\quad\mbox{for }(i,j,k)=(1,2,3)\]
where $n_{1}$, $n_{2}$, and $n_{3}$ are the densities of each kind
of atoms. The variation of the measured loss rate coefficient with
respect to the intensity $B$ of the applied magnetic field is shown
in the bottom panel of Fig.~\ref{fig:Panels}. It reveals a peak
around $B=$130~G which suggests an enhancement due to a resonance
of three colliding atoms with a three-body bound state. We can envisage
two kinds of resonance, depending on the origin of such a three-body
bound state.

Firstly, the three-body bound state may originate from another hyperfine
channel and couple by hyperfine interaction to the three-body scattering
continuum in the hyperfine channel $\vert1\rangle\vert2\rangle\vert3\rangle$.
Since the bound state and scattering state belong to different hyperfine
channels, they have different magnetic moments and become resonant
only around a particular intensity of the magnetic field which brings
them to the same energy. This situation would correspond to a \textquotedbl{}three-body
Feshbach resonance\textquotedbl{} \cite{rf:mehta}, a generalisation
of the now well-known two-body Feshbach resonances which occur for
two scattering atoms around certain magnetic field values \cite{rf:KohlerReview}.

As a matter of fact, wide two-body Feshbach resonances are present
in lithium 6 over the range of magnetic field intensities where the
3-body losses are observed. As a result, the two-body scattering lengths
between two atoms in different states, namely $\vert1\rangle\vert2\rangle$,
$\vert1\rangle\vert3\rangle$, and $\vert2\rangle\vert3\rangle$,
are modified by the applied magnetic field. The dependence of these
three scattering lengths on the magnetic field intensity is shown
in the top panel of Fig.~\ref{fig:Panels}. Because of this dependence,
the interactions between three atoms colliding in the hyperfine channel
$\vert1\rangle\vert2\rangle\vert3\rangle$ are also modified by the
magnetic field, and it may happen that a 3-body bound state supported
by these interactions within the same hyperfine channel is brought
to the threshold of its three-body scattering continuum at a certain
magnetic field value, causing a \textquotedbl{}shape resonance\textquotedbl{}.
This constitutes the second possible kind of resonance.

Studying both kinds of resonance theoretically is involved and requires
an extremely accurate knowledge of the interactions between atoms.
However, in the case where the two-body scattering lengths are much
larger than the range of the interatomic interactions, it is possible
to predict the structure of the three-body bound states near threshold
and their \textquotedbl{}shape resonance\textquotedbl{} simply in
terms of the scattering lengths and a short-range 3-body parameter.
This was pointed out by V.~N.~Efimov \cite{rf:efimov}, and the
corresponding three-body bound states, known as \textquotedbl{}Efimov
trimers\textquotedbl{}, are purely quantum-mechanical states which
enjoy special properties such as discrete scale invariance as the
scattering lengths are varied. In particular, their energy spectrum
forms an infinite series with a point of accumulation just below the
continuum threshold when the scattering lengths become infinite. So
far, signatures of Efimov trimers of identical bosons have been observed
in helium~4 \cite{rf:schollkopf}, cesium~133 \cite{rf:kraemer,rf:knoop},
and potassium 39 \cite{rf:zaccanti} and have been assigned in the
first two cases to the ground state of the Efimov series \cite{rf:lee,rf:braaten},
and to the ground and first excited state in the case of potassium.
Similar signatures of heteronuclear bosonic Efimov trimers of potassium
and rubidium were recently reported in Ref.~\cite{rf:barontini},
while the evidence of fermionic Efimov trimers is yet to be found.
It was suggested by the authors of Refs.~\cite{rf:ottenstein,rf:huckans}
that their observations in three-component lithium 6 might be the
manifestation of an Efimov trimer of distinguishable fermions. The
purpose of this Letter is to test this hypothesis using the Efimov
theory with the parameters of the experiments.

First, it should be noted that the two-body scattering lengths in
the experimental conditions are indeed quite large, from about -100
to -1000 $a_{0}$ (where $a_{0}=5.292\;10^{-11}$ m is the Bohr radius),
but not always much larger than the range of the atomic interactions,
typically given by the van der Waals length $\ell_{vdW}=\left(mC_{6}/\hbar\right)^{1/4}\approx60$
$a_{0}$, where $C_{6}$ is the van der Waals dispersive coefficient,
$m$ is the mass of lithium 6, and $\hbar$ is the reduced Planck's
constant. Therefore, the applicability of Efimov theory is questionable,
especially at low (around 100 G) and high (around 500 G) magnetic
field values where one or two of the scattering lengths become small.
However, in the intermediate region, the necessary conditions for
the existence of Efimov trimers are met.

The details of the Efimov theory can be found in Refs.~\cite{rf:efimov,rf:braaten}.
It essentially treats the free three-body problem with boundary conditions
at short distance imposing the known two-body scattering lengths between
each pair of atoms. The three-body wavefunction $\Psi(R,\alpha,\theta)$
is expressed in terms of the hyperradius $ $$R$ of the three-body
system (a measure of the global distance between the three atoms)
and the two angles $\alpha$ and $\theta$ describing the geometrical
configuration of the atoms. More precisely, if we denote by $\vec{r}$
the relative vector between atom 1 and 2, and by $\vec{\rho}$ the
relative vector between atom 3 and the centre of mass of atoms 1 and
2, then $R^{2}=r^{2}+\frac{4}{3}\rho^{2}$, $\theta$ is the angle
between $\vec{r}$ and $\vec{\rho}$, and $\mbox{tan}\alpha=\sqrt{3}r/(2\rho)$.
Using the Faddeev decomposition \cite{rf:faddeev}, and restricting
ourselves to the case of zero total angular momentum which is the
most favourable for the Efimov effect to occur \cite{rf:efimov},
the wave function can be written as\[
\Psi(R,\alpha,\theta)=\frac{2}{R^{2}}\Big(\frac{\tilde{\chi}^{(1)}(R,\alpha_{+})}{\sin2\alpha_{+}}+\frac{\tilde{\chi}^{(2)}(R,\alpha_{-})}{\sin2\alpha_{-}}+\frac{\tilde{\chi}^{(3)}(R,\alpha)}{\sin2\alpha}\Big),\]
where $\mbox{sin}2\alpha_{\pm}=\big[1-\big(\frac{1}{2}\cos\alpha\pm\frac{\sqrt{3}}{2}\sin\alpha\cos\theta\big)^{2}\big]^{1/2}$.

Note that the functions $\tilde{\chi}^{(i)}$ depend on only one hyperangle,
because we assumed that the total angular momentum is zero. They satisfy
the free Schrödinger equations\begin{equation}
\Big(\frac{\partial^{2}}{\partial R^{2}}+\frac{1}{R}\frac{\partial}{\partial R}+\frac{1}{R^{2}}\frac{\partial^{2}}{\partial\alpha^{2}}+\frac{mE}{\hbar^{2}}\Big)\tilde{\chi}^{(i)}(R,\alpha)=0\label{eq:Equations}\end{equation}
for a given energy $E$, with the boundary conditions\begin{multline}
\!\!\!\!\!\!\!\!\!\!\!\!\!\frac{\partial\tilde{\chi}^{(i)}}{\partial\alpha}(R,0)+\frac{4}{\sqrt{3}}\big(\tilde{\chi}^{(j)}(R,\!\!\!\begin{array}{c}
\frac{\pi}{3}\end{array}\!\!\!)+\tilde{\chi}^{(k)}(R,\!\!\!\begin{array}{c}
\frac{\pi}{3}\end{array}\!\!\!)\big)=-\frac{R}{a_{jk}}\tilde{\chi}^{(i)}(R,0),\label{eq:Condition1}\end{multline}
\begin{equation}
\tilde{\chi}^{(i)}(R,\!\!\!\begin{array}{c}
\frac{\pi}{2}\end{array}\!\!\!)=0\label{eq:Condition2}\end{equation}
for any $R>R_{0}$, and\begin{equation}
\frac{\partial\ln\tilde{\chi}^{(i)}}{\partial R}(R_{0},\alpha)=\Lambda(R_{0}),\quad\mbox{for any }\alpha\in[0,\!\!\!\begin{array}{c}
\frac{\pi}{2}\end{array}\!\!]\label{eq:Condition3}\end{equation}
where $i,j,k$ is any permutation of $1,2,3$.

The first boundary condition (\ref{eq:Condition1}) imposes the form
of the wave function in the two-body sectors consistent with the known
two-body scattering lengths $a_{12}$, $a_{13}$, and $a_{23}$ between
the three kinds of atoms. The last boundary condition (\ref{eq:Condition3})
fixes the logarithmic derivative of the wave function at short hyperradius
$R_{0}\ll a_{ij}$ to some value $\Lambda(R_{0})$ independent of
the energy $E$. In this region, the last term of Eq.~(\ref{eq:Condition1})
is negligible, and one can show that the solution of Eqs.~(\ref{eq:Equations})
takes the separable form $\tilde{\chi}^{(i)}(R,\alpha)\approx\sin(\vert s_{0}\vert\ln(KR)+\Delta)\sin(s_{0}(\pi/2-\alpha))$,
where $K=\sqrt{mE}/\hbar$, $\Delta$ is a phase shift, and $s_{0}\approx1.00624i$
is the imaginary solution of the equation $-s_{0}\cos(s_{0}\pi/2)+8/\sqrt{3}\sin(s_{0}\pi/6)=0$,
which follows from applying the boundary condition Eq.~(\ref{eq:Condition1}).
In order for $\Lambda$ to be energy independent, the phase shift
$\Delta$ must be of the form $-\vert s_{0}\vert\ln(K/\Lambda_{0})$
where $\Lambda_{0}$ is some fixed wave number. This sets \[
\Lambda(R_{0})=\frac{\vert s_{0}\vert}{R_{0}}\cot\left[\vert s_{0}\vert\log(R_{0}\Lambda_{0})\right].\]
This way, the choice of $R_{0}$ is arbitrary. The only requirement
is that $R_{0}$ should be much smaller than the scattering lengths.
In our calculation, we fixed $R_{0}$ to 1 $a_{0}$. Thus, the only
free parameter of the theory is $\Lambda_{0}$. It captures the effects
of the unknown short-range three-body physics on the wave function
at larger hyperradii.

We solve these equations by discretising the arguments $(R,\alpha)$
of the functions $\tilde{\chi}^{(i)}$ on a 2-dimensional grid, evaluating
the derivatives by finite differences, and diagonalising the resulting
matrix corresponding to the Hamiltonian of the system. The maximal
value of $R$ is set to 50,000~$a_{0}$, which is on the order of
the particle spacing in the experiments. At each value of the magnetic
field, the three scattering lengths $a_{12}$, $a_{13}$, and $a_{23}$
are obtained from the top panel of Fig.~\ref{fig:Panels}. The only
unknown quantity is $\Lambda_{0}$. First, we make the assumption
that it does not depend on the magnetic field. In reality, it may
actually depend on it, but provided that no accidental resonance occurs,
it is reasonable to assume that its variations are less pronounced
than that of the scattering lengths, at least in first approximation.
Second, we make the fundamental assumption that the observed three-body
losses are due to a shape resonance with an Efimov trimer. The measured
three-body loss rate coefficient as a function of the intensity of
the magnetic field shows a distinctive peak around $B=$130~G - see
the bottom panel of Fig.~\ref{fig:Panels}. Assuming that this is
the point where an Efimov state reaches the continuum threshold (its
binding energy goes to zero), we find that we should adjust $\Lambda_{0}$
to about $(0.42\; a_{0})^{-1}$. Once $\Lambda_{0}$ is fixed, we
can obtain the eigenstates and their energy around the threshold,
namely the Efimov trimers and the three-body scattering states. 

In the middle panel of Fig.~\ref{fig:Panels} we plotted the energy
of the Efimov trimer just below the continuum threshold as a function
of magnetic field. By construction, the trimer appears at $B=130$~G.
Interestingly, its binding energy increases until $B=350$~G, and
then decreases until the trimer reaches the continuum again, causing
a second zero-energy resonance around $B=500$~G. This simply results
from the magnetic field dependence of the scattering lengths.

\begin{figure}
\includegraphics{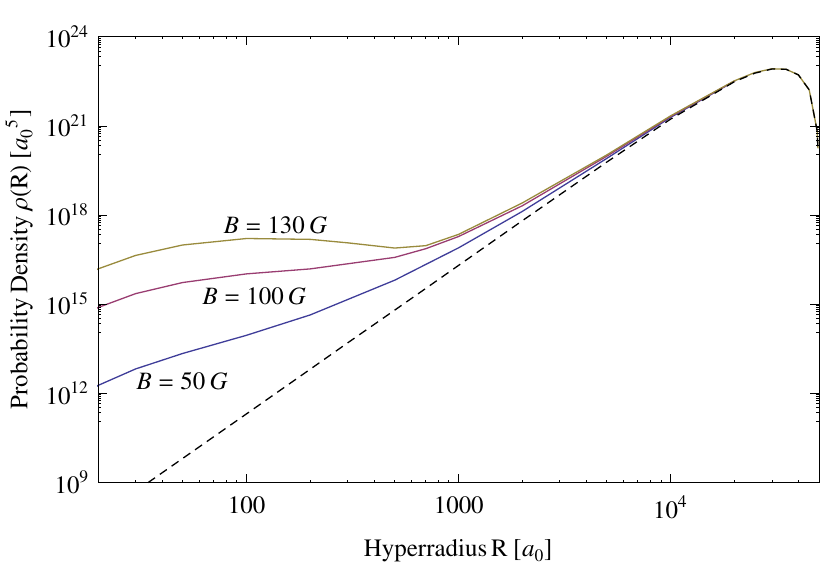}

\caption{\label{fig:HyperradialProb}Probability density $\rho(R)$ - defined
by Eq.~(\ref{eq:HyperradialProb}) - of the lowest three-body scattering
state for different values of the magnetic field. In all cases, the
wave function $\Psi$ is normalised to be asymptotically equal to
the noninteracting limit $8J_{2}(KR)/(KR)^{2}$, where $J_{2}$ is
the Bessel function, and $K$ is the wave number. This limit is indicated
by the dashed curve. Here, the wave number $K$ is set by the size
of the numerical grid.}

\end{figure}

Since the two-body scattering lengths are all negative in this range
of magnetic field, there is no two-body bound states just below the
two-body continuum. As a result, when inelastic three-body collisions
occur, two of the three atoms have to form a deeply bound dimer. A
direct calculation of the rate for such processes would require a
detailed analysis of the deeply bound dimers. However, we can easily
calculate its enhancement by the Efimov resonance. Indeed, by assigning
a complex value $\vert\Lambda_{0}\vert e^{i\eta}$ to the 3-body parameter
$\Lambda_{0}$, and therefore a complex value to the logarithmic derivative
$\Lambda$, one can impose a probability loss at short hyperradius
($R=R_{0}$) in order to model the overall effect of losses by recombination
to deeply bound dimers \cite{rf:braaten}. This makes any scattering
state $\Psi$ quasistationary, and by calculating the time variation
of the total probability $\int\vert\Psi\vert^{2}d^{3}\vec{R}d^{3}\vec{r}$,
one can derive the following 3-body loss rate coefficient:\begin{equation}
\mathcal{K}=\frac{2\hbar}{m}\vert\mbox{Im}\,\Lambda\vert\;\times\;\rho(R_{0}).\label{eq:RateCoefficient}\end{equation}
As one would expect, this coefficient is proportional to the imposed
velocity $\mathcal{V}=\frac{2\hbar}{m}\vert\mbox{Im}\,\Lambda\vert$
at $R=R_{0}$ and the probability density $\rho(R_{0})$ of finding
three atoms at that hyperradius, defined by the hyperangular average\begin{equation}
\rho(R)=3^{\frac{3}{2}}\pi^{2}R^{5}\!\!\int_{0}^{\pi/2}\!\!\!\!\!(\!\!\begin{array}{c}
\frac{1}{2}\end{array}\!\!\sin2\alpha)^{2}d\alpha\!\!\int_{0}^{\pi}\!\!\!\sin\theta d\theta\vert\Psi(R,\alpha,\theta)\vert^{2},\label{eq:HyperradialProb}\end{equation}
where $\Psi$ is normalised to be asymptotically equal to the noninteracting
limit $8J_{2}(KR)/(KR)^{2}$, $J_{2}$ being the Bessel function.
This probability density, and thus the inelastic processes, is strongly
increased at short distance by the presence of an Efimov trimer just
below threshold. Physically, this is due to the fact that the three
atoms almost bind during their collision, and therefore spend more
time together. One can see in Fig.~\ref{fig:HyperradialProb} that
while $\rho(R)$ is unaffected at large distance, it changes significantly
at short distance when the magnetic field is varied around the zero-energy
resonance.

The calculated loss rate coefficient of Eq.~(\ref{eq:RateCoefficient})
is plotted in the bottom panel of Fig.~\ref{fig:Panels}. We obtain
a profile delimited by two peaks. In between the two peaks, we observe
a plateau due to the presence of the Efimov state below threshold.
Outside, the probability becomes very low, due to the absence of a
near-threshold Efimov trimer. From the energy spectrum, it is clear
that the second peak around 500 G is related to the second zero-energy
resonance with the Efimov state. We can then adjust the value of $\eta$
to fit the experimental data. It should be noted that $\eta$ has
two effects: it sets the overall magnitude of the loss rate coefficient,
as can be seen from (\ref{eq:RateCoefficient}), and it also smoothes
the peaks (because they are naturally broadened by the losses). It
turns out that there exists a range of values for $\eta$ which can
approximately fit both the overall magnitude of the rate coefficient
and its shape. Using a least-square minimization method, we found
that we should set $\eta$ to 0.157 in order to obtain the best fit
to the experimental data. Fitting only the shape of the first peak
with that of the experimental peak at $B=130$ G, we obtain the best
agreement for $\eta=0.115$.

The behaviour of the calculated three-body decay rate coefficient
is very reminiscent of the measured one, which has a similar profile
between 130 and 500 G. This suggests that the local maximum near $B=$
500~G found in the experiment is caused by a second resonance. The
agreement with the observations is only approximate however, as the
experimental data show a much more diffuse local maximum. As we noted
earlier, the Efimov theory is not strictly applicable to the present
system, and it is expected that short-range corrections are needed
for a better agreement. A magnetic-field dependence of $\Lambda_{0}$
might also play a role. These effects may very well explain the remaining
discrepancies. Yet, it is quite remarkable that the Efimov theory
already provides a basic description which qualitatively explains
the experimental observations. In this interpretation, a relatively
\textquotedbl{}pure\textquotedbl{} Efimov trimer state is expected
around 300~G (where the scattering lengths are largest) and connects
continuously to trimer states which are partly affected by short-range
physics (\textquotedbl{}impure\textquotedbl{} Efimov trimer) and eventually
dissociate in the continuum at both resonances. We find that the pure
Efimov trimer at $B=$300~G has a binding energy of about $2\pi\hbar\times$10~MHz
and a width (imaginary part of the energy) of about $2\pi\hbar\times$3~MHz,
corresponding to a lifetime of about 50~ns. However, these values
result from an adjustment near $B=130$~G, where the Efimov theory
may need corrections. Thus, the actual values might be slightly different.

The present interpretation may be confirmed or refuted by further
experimental investigation, in particular direct observation of the
trimers below threshold - for example, by radiofrequency spectroscopy.
For comparison with the first interpretation discussed in the introduction
(Feshbach resonance with another hyperfine channel), we calculated
the expected energy of possible resonant trimers coming from other
channels, assuming that their magnetic moment is simply the sum of
the magnetic moments of the three separated atoms. Since the interaction
conserves the projection of the total spin, we considered only the
channels with the same projection $m_{F}=-3/2$ as the incoming channel
$\vert1\rangle\vert2\rangle\vert3\rangle$, and shifted the energy
of the possible resonant trimers such that it crosses the 3-body threshold
of the incoming channel at $B=130$~G. The resulting energies as
a function of magnetic field are plotted in Fig.~\ref{fig:Panels}
as dashed curves. One can see that they have a monotonic behaviour,
and depart steeply from the continnum threshold. On the other hand,
if the Efimov interpretation is correct, the trimer is expected to
follow a different and rather unusual behaviour: it connects to the
continuum via two zero-energy resonances, resulting in an energy minimum
as function of magnetic field. Thus we have provided a significant
difference between the two scenarios, which may serve as a test in
future experiments.

We are grateful to S.~Jochim for providing the experimental data
shown in Fig.~1.\\

\textit{Note added}: Upon finishing this Letter, we became aware of
the work by E.~Braaten et al. \cite{rf:braaten2}, which uses the
Skorniakov\textendash{}Ter-Martirosian equations to directly calculate
the loss rate from scattering amplitudes. After this Letter was submitted,
similar results were reported in Ref.~\cite{rf:floerchinger}, using
a different model involving the formation of a trimer. Both of these
works yield results which are consistent with the loss rate coefficient
shown in Fig.~1.

\end{document}